\documentclass[aps,pre,twocolumn,showpacs,10pt,longbibliography]{revtex4-1}
\usepackage{graphicx}
\usepackage{epstopdf}
\usepackage{amsmath}
\usepackage{amssymb}
\usepackage{default}
\usepackage{default}
\newcommand{\be}{\begin{equation}}
\newcommand{\bea}{\begin{eqnarray}}
\newcommand{\bc}{\begin{center}}            
\newcommand{\ee}{\end{equation}}
\newcommand{\eea}{\end{eqnarray}}
\newcommand{\ec}{\end{center}}

\newcommand{\baa}{\begin{eqnarray*}}
\newcommand{\eaa}{\end{eqnarray*}}
\begin{document}
\title{Heat engines at optimal power: \\ 
Low-dissipation versus endoreversible model
}
\author{Ramandeep S. Johal}
\email{rsjohal@iisermohali.ac.in}
\affiliation{Department of Physical Sciences, \\ 
Indian Institute of Science Education and Research Mohali,\\  
Sector 81, S.A.S. Nagar, Manauli PO 140306, Punjab, India}
\begin{abstract}
Low-dissipation model and the endoreversible model
of heat engines are two of the most commonly studied
models of machines in finite-time thermodynamics. In this paper,
we compare the performance characteristics of these
two models under optimal power output. We point out
a basic equivalence between them, in the linear response regime.
\end{abstract}
\maketitle
\section{Introduction}
The ideal heat cycles discussed in
text books, are incapable of modelling  
realistic thermodynamic machines. For instance, Carnot engine yields 
maximum work, but its power output (defined as work output per cycle time) is
practically zero---due to infeasibly large cycle times. 
Secondly, the ideal cycle involves no net entropy change in the environment,
whereas the operation of real machines, always entails a positive entropy
generation. 
In recent years, research on finite-time models of thermodynamic machines
has gained a lot of attention \cite{Berry1984, Bejan1996, Salamon2001, Andresen2011}. 
Irreversibilities can be incorporated by assuming a finite rate for 
heat transfer, internal friction, and heat leakage. 
Models based on linear irreversible thermodynamics 
\cite{Broeck2005, Wangtu2012}, 
the assumption of endoreversibility \cite{Curzon1975, Chen1989, Apertet2013, Correa2014},
and weak or low dissipation 
\cite{Schmiedl2008, Esposito2010, Roco2012,Roco2012b, Broeck2013,Holubec2016,Roco2017}, 
are some of the approaches which have been pursued.
On the other hand, the finite size of reservoirs, in contrast to
infinite reservoirs, also reduces the performance  
\cite{Ondrechen1981, Leff1987,Izumida2014,JohalRai2016}.  

In this paper, we focus on the characteristics---at optimal power output---of
two currently studied models in finite-time thermodynamics, viz.
low-dissipation model and the endoreversible model. 
We highlight the similarities and differences between the two models.
In particular, we show their equivalence in the linear response regime, i.e.
for small difference of bath temperatures.

The plan of the paper is as follows. In Section II, we discuss
the optimal features of low-dissipation model. In Section III, 
we describe optimal features of an endoreversible model assuming
a linear irreversible law for heat transfer. In Section IV, we
compare the two models. 


\section{Low-dissipation model: Optimal operation}
Consider a two heat-reservoirs setup, with hot ($h$) and cold 
($c$) temperatures, $T_h$ and $T_c$. A heat engine runs through a four-step 
cycle by coupling to these reservoirs alternately.
The cycle consists of two thermal contacts lasting for 
time intervals $\tau_h$ and $\tau_c$, and two adiabatic steps whose time 
intervals may be neglected in comparison to the other time scales.
Now, the change in entropy of the working medium 
during heat transfer at the hot/cold  
contact, can be split as: $\Delta S_{j} = \Delta_{\rm rev} S_{j} + \Delta_{\rm ir} S_{j}$,
with $j=c,h$. Here, the first term accounts for a reversible 
heat transfer (equal to the amount of heat transferred, 
divided by the temperature of the 
reservoir), whereas the second term denotes an irreversible entropy 
generation during the process. Now, 
the low-dissipation assumption, which is expected to apply close to
the reversibility limit, models the latter term 
as being inversely proportional to the duration of the time spent ($\tau_{c,h}$)
on the heat transfer step: 
$T_j \Delta_{\rm ir} S_{j} = \sigma_j/ \tau_j + O(1/\tau_{j}^{2})$,
where $\sigma_{j}$ is the dissipation constant \cite{Esposito2010,Broeck2013}. Thus
at the hot and the cold contact, we respectively have
\bea
\Delta S_h &=& \frac{Q_h}{T_h} + \frac{\sigma_h}{T_h \tau_h},
\label{dsh} \\
\Delta S_c &=& -\frac{Q_c}{T_c} + \frac{\sigma_c}{T_c \tau_c},
\label{dsc}
\eea
where $Q_j > 0$. Given that the other two steps in the heat cycle are 
adiabatic---with no entropy changes---the cyclic process within the working medium implies 
$\Delta S_h + \Delta S_c = 0$. In other words, 
$\Delta S_h = -\Delta S_c = \Delta S > 0$,
where the value $\Delta S$ is preassigned.
Then the amount of heat exchanged with each reservoir can be written as:
\bea
Q_h &=& T_h \Delta S - \frac{\sigma_h}{\tau_h}, 
\label{qh} \\
Q_c &=& T_c  \Delta S + \frac{\sigma_c}{\tau_c}. 
\label{qc}
\eea
The work extracted in a cycle with the time period $\tau \approx \tau_h + \tau_c$ is,  
$W = Q_h - Q_c$. So the average power output per cycle is defined as 
\be
P = \frac{Q_h - Q_c}{\tau_h + \tau_c}.
\label{pwr}
\ee
Although, the working medium undergoes a cyclic process--with no net
entropy change, the whole cycle is irreversible, with a net total 
entropy production per cycle. This is given by the net change in 
the entropies of the two reservoirs: 
\be 
\Delta _{\rm tot} S = -\frac{Q_h}{T_h} + \frac{Q_c}{T_c}.
\ee
Now, in finite-time thermodynamics, quite often, the desired 
objective is to maximize the power output. 
Thus, in order to determine the operating conditions for the maximum  
power, we set
\be
\left(\frac{\partial P}{\partial \tau_h }\right)_{\tau_c}  = 0,
\quad  \left( \frac{\partial P}{\partial \tau_c } \right)_{\tau_h} = 0,
\label{dphc}
\ee
where we assume certain given values of the parameters $\sigma_h, \sigma_c$ and $\Delta S$.
This yields the optimal allocation for the contact times, given by:
\bea
\hat{\tau}_{h} &=& \frac{2}{\Delta T \Delta S }
\sqrt{\sigma_h} (\sqrt{\sigma_h} + \sqrt{\sigma_c} ), 
\label{th} \\
\hat{\tau}_{c} &=& \frac{2}{\Delta T \Delta S }
\sqrt{\sigma_c} (\sqrt{\sigma_h} + \sqrt{\sigma_c} ),
\label{tc}
\eea
where $\Delta T = T_h - T_c$. 
Substituting Eq. (\ref{th}) in (\ref{qh}),
and Eq. (\ref{tc}) in (\ref{qc}), we obtain explicit 
expressions (at optimal power) for the amounts of heat 
transferred at each thermal contact:
\bea
\hat{Q}_{h} &=& T_h \Delta S -  \frac{\gamma}{2} {\Delta T \Delta S},  
\label{qhstar} \\
%
\hat{Q}_{c}   & = & T_h \Delta S -  \frac{(1+\gamma)}{2}{\Delta T \Delta S}, 
\label{qcstar}
\eea
where $\gamma = (1+ \sqrt{\sigma_c /\sigma_h})^{-1}$.
From Eqs. (\ref{qhstar}) and (\ref{qcstar}), 
 the work output per cycle is given by
\be
\hat{W} = \hat{Q}_{h} - \hat{Q}_{c} = \frac{1}{2} \Delta T \Delta S.
\label{wstar}
\ee
The efficiency at optimal power,
defined as $\hat{\eta} =\hat{W}/\hat{Q}_h$, takes the following form:
\be
\hat{\eta} = \frac{\eta_C}{2-\gamma \eta_C}, 
\label{etg}
\ee
where $\eta_C = 1-T_c/T_h$, is the Carnot value.
Thus the efficiency at optimal power 
is bounded as:
\be
\frac{\eta_C}{2} \leqslant \hat{\eta} \leqslant \frac{\eta_C}{2 - \eta_C}.
\label{bsef}
\ee
Then the following extreme cases are of interest.
When $\sigma_c  \ll \sigma_h$, or 
$\gamma \to 1$, it means that the heat transfer at the cold
contact approaches the reversible limit, and the efficiency
approaches the upper bound.
On the other hand, under the condition $\sigma_h  \ll \sigma_c$, or 
$\gamma \to 0$, the hot contact approaches
the reversible limit, and the efficiency approaches its lower bound.
\subsection{Rates of Dissipation}
We also note that, under optimal power conditions,
the amounts of dissipation at the hot and the cold  
contacts, defined by $T_j \Delta_{\rm ir} \hat{S}_j$,
is respectively given by:
\be
T_h \Delta_{\rm ir} \hat{S}_h = \frac{\sigma_h}{\hat{\tau}_h} 
    =   \frac{\gamma}{2} \Delta T \Delta S, \label{dish}
   \ee
and
 \be
T_c \Delta_{\rm ir} \hat{S}_c = \frac{\sigma_c}{\hat{\tau}_c} 
   =   \frac{1-\gamma}{2} \Delta T \Delta S. \label{disc}
   \ee   
   Then, at optimal power,
   the average rates of dissipation at the two thermal 
   contacts, are equal:
   \be
   \frac{T_h \Delta_{\rm ir} \hat{S}_h}{\hat{\tau}_h}
   = \frac{T_c \Delta_{\rm ir} \hat{S}_c}{\hat{\tau}_c} =
   \left[ \frac{\Delta T \Delta S }{2(\sqrt{\sigma_h} + \sqrt{\sigma_c})} \right]^2.
   \ee
   Incidentally, the above rate of dissipation is same as the 
   optimal power output, $\hat{P}$.  
   
   \section{Endoreversible model with linear irreversible law}
   In the so-called endoreversible models \cite{Curzon1975,Andresen2011},
   a specific form of heat-transfer law is assumed between a reservoir 
   and the working medium \cite{Chen1989}. Basically, irreversibility 
   arises due to flow of heat---with a finite rate---across
   a finite heat conductance. In the following, we consider
   such a model where the heat flux is proportional to the difference
   of inverse temperatures of the working medium and the reservoir.
   This particular law is based on the 
   flux-force relation in linear irreversible thermodynamics \cite{Grootbook},
   and is applicable for small temperature gradients.
   For brevity, we address this model as the linear model. 
   Now, consider 
   $T_1$ and $T_2$ to be the {\it fixed} temperatures of the working medium 
   at hot and cold contacts respectively. 
   Then the heat fluxes are given by 
\bea 
q_h & = & \alpha_h \left ( T_{1}^{-1} - T_{h}^{-1}      \right ), \label{rateh}\\
q_c & = & \alpha_c \left ( T_{c}^{-1} - T_{2}^{-1}      \right ), \label{ratec}
   \eea
where $\alpha_j$, with $j=c,h$ be the heat conductance.
As the fluxes are constant during the times of contact, 
so the amounts of heat transferred during the times 
$t_h$ and $t_c$, respectively are:
$Q_h = q_h t_h$ and $Q_c = q_c t_c$.
%
The entropy change in the working medium at hot and 
cold contacts will be: $\Delta' S_h = Q_h/T_1$ and $\Delta' S_c = -Q_c/T_2$,
respectively. Again, in the adiabatic steps, the
entropy of the working medium stays constant. 
The cyclicity within the 
working medium implies, $\Delta' S_h  = -\Delta' S_c = \Delta' S >0$,
which yields 
\be
\frac{Q_h}{T_1} = \frac{Q_c}{T_2} \equiv \Delta' S,
\label{endo}
\ee
which is usually known as the endoreversibility condition.
Again, the work extracted per cycle is $W = Q_h-Q_c$,
and the average power per cycle is $P = (Q_h - Q_c)/(t_h + t_c)$.
The efficiency per cycle is
\be
\eta = 1-\frac{Q_c}{Q_h} = 1-\frac{T_2}{T_1}.
\label{effl}
\ee   
Now, we  optimize the power with respect to variables 
$T_1$ and $T_2$. The maximum power is obtained at the following
values:
\bea
\tilde{T}_1 &=& \frac{2(1-\eta_C)}{2-(1+\bar{\gamma})\eta_C} T_h , \label{optt1} \\
\tilde{T}_2 & =& \frac{2 T_c}{2-\bar{\gamma}\eta_C}.
\label{optt2}
\eea
The efficiency at maximum power is
\be
\tilde{\eta} = \frac{\eta_C}{2-\bar{\gamma} \eta_C}, 
\label{etgp}
\ee
where $\bar{\gamma} = (1+ \sqrt{\alpha_c/\alpha_h})^{-1}$.

Using Eqs. (\ref{optt1}) and (\ref{optt2}),
the heat fluxes at optimal power conditions are given by:
\bea 
\tilde{q}_h & = & \frac{1}{2}{\alpha_h (1-\bar{\gamma})}
\left ( {T}_{c}^{-1} - T_{h}^{-1}      \right )  \label{ratehopf}\\
\tilde{q}_c & = & \frac{1}{2}{\alpha_c \bar{\gamma}}
\left ( {T}_{c}^{-1} - T_{h}^{-1}      \right )  \label{ratecopf},
\eea
so that $\tilde{q}_h/\tilde{q}_c = \sqrt{\alpha_h/ \alpha_c }$.

From Eq. (\ref{endo}), we can write for optimal power conditions:
\bea 
\tilde{Q}_h &=&   \frac{2(1-\eta_C)}{2-(1+\bar{\gamma})\eta_C} T_h  \Delta' S,
\label{heathil} \\
\tilde{Q}_c &=&   \frac{2(1-\eta_C)}{2- \bar{\gamma} \eta_C} T_h  \Delta' S,
\label{heatcil}
\eea
from which the extracted work per cycle is given by: 
$\tilde{W} = \tilde{Q}_h - \tilde{Q}_c$.
From the above expressions, one can easily see that 
for $\bar{\gamma} \to 1$, the hot contact approaches
reversible limit, and the efficiency at maximum power 
approaches its upper bound of $\eta_C /(2-\eta_C)$.
Similarly, for $\bar{\gamma} \to 0$,
the cold contact becomes reversible and the efficiency
approaches the lower bound of $\eta_C /2$.

\subsection{Entropy generation}
Let us consider the entropy generated at each thermal contact.
For the hot contact:
\be
\Delta_{\rm ir} S_h = Q_h \left ( T_{1}^{-1} - T_{h}^{-1}      \right ).
\label{irsh}
\ee
The average rate of entropy generation at the hot contact is:
\bea 
\frac{\Delta_{\rm ir} S_h}{t_h} &=& \frac{Q_h}{t_h} 
\left ( T_{1}^{-1} - T_{h}^{-1}      \right )  \\
& = & \frac{q_{h}^{2}}{\alpha_h}.
\eea
At optimal power, the above rate of entropy generation is given by:
\be
\frac{\Delta_{\rm ir} \tilde{S}_h}{t_h} =  \frac{ {\tilde{q}_{h}}^2  }{\alpha_h},
\ee
see Eq. (\ref{ratehopf}). 
Similarly, we have the corresponding expression
of entropy generation, at the cold contact,  
$\Delta_{\rm ir} \tilde{S}_c = 
\tilde{Q}_c ( T_{c}^{-1} - \tilde{T}_{2}^{-1})$   
with corresponding rate of entropy generation:
\be
\frac{\Delta_{\rm ir} \tilde{S}_c}{t_c} =  \frac{ {\tilde{q}_{c}}^{2}  }{\alpha_c}.
\ee
Using the expressions for heat fluxes at optimal power, Eqs. (\ref{ratehopf})
and (\ref{ratecopf}), we reach the conclusion that, in the endoreversible
model, the rates of entropy
production are equal at the hot and the cold contacts, under optimal power. 
\section{The comparison}
%
Although the two models are based on seemingly different
assumptions, there is a remarkable 
similarity between the expressions for efficiency, 
(\ref{etg}) and (\ref{etgp}), at optimal power. 
It is apparent that
the parameters $\Delta S$ and $\Delta' S$ also 
play analogous roles in these models. 
However, there are points of difference.
Thus the expressions for the heat exchanged
with the reservoirs, and the work performed per cycle
appear to be different. Secondly, as has been shown
above, in low-dissipation model, the rates of dissipation
at the hot and cold contacts become equal, whereas
in endoreversible model, it is the two rates of entropy
generation, that are equal at optimal power. 
Also in a sense, the parameters $\gamma$ and 
$\bar{\gamma}$ play complementary roles. Thus 
$\gamma \to 1$ corresponds to $\bar{\gamma} \to 0$,
which is understandable since if the dissipation
constant at the cold contact becomes vanishingly small,
it implies approach to reversible limit at that contact.
The analogous condition for endoreversible model
is that the conductance at the cold contact becomes 
very large. Similarly, we expect that 
$\gamma \to 0$ corresponds to $\bar{\gamma} \to 1$.

However, as we show below, if we identify the 
common domain of validity for these models, 
then the optimal performance of these
apparently different models exhibits a basic equivalence. 

In fact, the linear irreversible law (Eqs. (\ref{rateh}) and (\ref{ratec})),
is expected to be applicable for small temperature differences. 
For the hot contact, it implies that 
$1-T_1/T_h \ll 1$. Under conditions of optimal power, this condition 
is $1-\tilde{T}_1/T_h \ll 1$, which from
Eq. (\ref{optt1}) gives the condition $\eta_C \ll 1$.
Applying a similar argument to the cold contact---at optimal power,
the corresponding condition 
is given by $\tilde{T}_2 /T_c -1 \ll 1$,
which yields $\bar{\gamma}\eta_C \ll 1$.
Since $\bar{\gamma}$ lies between zero and unity, so 
the essential condition is, $\eta_C \ll 1$.
Thus we see that endoreversible model  at optimal power, with the linear
law, requires small temperature differences between
the reservoirs.

Thus in the linear response regime, which implies small
values of $\Delta T = T_h - T_c$,  
the expressions for heat in the endoreversible model,
(\ref{heathil}) and (\ref{heatcil}),
are simplified as follows:
\bea
\tilde{Q}_{h} &=& T_h \Delta' S -  \frac{1-\bar{\gamma}}{2} {\Delta T \Delta' S},  
\label{qhtil} \\
\tilde{Q}_{c} &=& T_h \Delta' S -  \frac{2-\bar{\gamma}}{2}{\Delta T \Delta' S}.
\label{qctil}
\eea
The above expressions may be compared to the corresponding
expressions (\ref{qhstar}) and (\ref{qcstar}) for the low-dissipation
model. The extracted work per cycle is: 
$\tilde{W} = \frac{1}{2}\Delta T \Delta' S.$
This shows that within linear response (upto first order in $\Delta T$), 
the corresponding expressions for heat and work extracted per cycle, 
are similar in both models. In particular, 
the parameter $\gamma$ is equivalent to $1-\bar{\gamma}$
in this limit. Accordingly, in this regime, the total entropy generated per cycle
in the environment, shows similar behavior within the two models.
 
The above comparison becomes interesting due 
to the fact that in the low-dissipation model, 
there is no intrinsic requirement
for the temperature difference $\Delta T$ to be small 
\cite{Esposito2010, Broeck2013}. 
However, as the comparison with the linear model 
shows, the expressions
at optimal power which are expected to hold in 
the linear response regime, also hold for arbitrary
temperature differences---according to the low dissipation model.
These observations indicate the need 
to analyze more thoroughy the domain of applicability
of the low-dissipation model. 

Concluding, we have clearly identified the points
of similarity, and difference, between the
low-dissipation and endoreversible model, under optimal power conditions.
The present study also identifies the 
equivalence of these two models within the linear response regime,
consistent with the principles of linear
irreversible thermodynamics.

%

\end{document}